# High-resolution Collinear Chiral Sum Frequency Generation Microscopy by Using Vectorial Beam

Ziheng Ji[1], Wentao Yu[1], Hong Yang[1], Kaihui Liu[1], Qihuang Gong[1,2], Zhiwen Liu[3] and Kebin Shi*[1,2]

[1]State Key Laboratory for Mesoscopic Physics, Collaborative Innovation Center of Quantum Matter, School of Physics, Peking University, Beijing 100871, China

[2]Collaborative Innovation Center of Extreme Optics, Shanxi University, Taiyuan, Shanxi 030006, China

[3]Department of Electrical Engineering, the Pennsylvania State University, University Park, PA 16802

*To whom correspondence should be addressed. E-mail: kebinshi@pku.edu.cn

**Abstract**: In chiral sum frequency generation (C-SFG), the chiral nature of $\overleftrightarrow{\chi}^{(2)}$ requires the three involved electric fields to be pairwise non-parallel, leading to the traditional non-collinear configuration which is a hindrance for achieving diffraction limited resolution while utilizing it as a label-free imaging contrast mechanism . Here we propose a collinear C-SFG (CC-SFG) microscopy modality by using longitudinal z-polarized vectorial field. Label-free chiral imaging with enhanced spatial resolution (~1.4 times improvement in one lateral and the longitudinal directions over the traditional non-collinear scheme) is demonstrated, providing a new path for SFG microscopy with diffraction-limited resolution for mapping chirality.

Chirality plays a vital role in both natural world and artificial realm. Most building blocks of life, amino acids, proteins, DNAs, are chiral [1]. Metamaterials with negative refractive index could be achieved by chiral metal structure [2]. Their functionalities on the mesoscopic scale are strongly correlated with their chiral structures. Characterizing chiral system has always been a challenge. Optical spectroscopy is one of the most powerful and efficient tools to probe chiral system under practical situation for noninvasive characterization. Optical rotatory dispersion (ORD) and circular dichroism (CD) have been widely applied, respectively, essentially by comparing differences in real and imaginary parts of refractive indices when illuminated with different handedness of light [3]. Their signals, generally, are rather weak since the process has its origin from magnetic-dipole and electric-quadrupole interaction [4]. Fluorescence detected circular dichroism (FDCD) is sensitive enough to probe the chirality of a single molecule [5], but in practice, this signal could be overwhelmed by a chirality-insensitive fluorescence background generated from the surrounding medium. Optically active Raman spectroscopy [6] is an alternative approach, and yet suffers from a low sensitivity. Recently, electronically resonant chiral sum-frequency generation (C-SFG) spectroscopy [7] has emerged as a promising technique to probe chirality free from achiral background, because it is only allowed in systems without inversion symmetry under the electric-dipole approximation. In addition, C-SFG could be enhanced by vibrational resonance thus providing chemical selectivity [8]. Although electronically resonant C-SFG requires a breakdown of the Born-Oppenheimer approximation, it still possesses the sensitivity to probe molecular monolayer [9].

Sum frequency generation (SFG) is a second order nonlinear optical process, which generates a signal at $\omega_{SF} = \omega_1 + \omega_2$ when two coherent laser beams with angular frequencies of $\omega_1, \omega_2$ interact in a medium possessing second order nonlinear susceptibility $\overleftrightarrow{\chi}^{(2)}$, which is a third order tensor. Note that chiral second harmonic generation is the frequency degenerate case of C-SFG, but is nevertheless unable to probe isotropic chiral system [7]. The nonlinear polarization source of the SFG signal is given by $\overleftrightarrow{\chi}^{(2)}:\vec{E}_1\vec{E}_2$. In isotropic chiral liquid, chiral molecules are randomly oriented thus leading to $\infty\infty$ symmetry. Any axis in the liquid is a $C_\infty$ axis, leaving only six nonzero elements of $\overleftrightarrow{\chi}^{(2)}$, namely, $\chi_{xyz}^{(2)} =$ $-\chi^{(2)}_{xzy} = \chi^{(2)}_{yzx} = -\chi^{(2)}_{yxz} = \chi^{(2)}_{zxy} = -\chi^{(2)}_{zyx}$, while all the other elements vanish in orientational averaging [10]. These six elements have three different vectorial indices. As a result, the three electric fields involved in the chiral SFG process must be pairwise non-parallel. For example, under the excitation of two orthogonally linearly polarized (LP) beams as shown in Figure 1 (a), the SFG cannot be effectively generated in the forward direction in the far field since the resultant second order nonlinear polarization $\vec{P}^{NL}$ is dominated by its longitudinal component. Consequently C-SFG is often performed in a non-collinear geometry by focusing two beams from distinct directions under SPP, PPS or PSP polarization configuration (P and S stand for in plane and out of plane polarization state, respectively; the three polarization states refer to that of the SFG, the first pump, and the second pump following the common convention in literature.) to satisfy the pair-wise non-parallel condition as shown in Figure 1 (b) [11,12].

Efforts have also been made to adapt the classical non-collinear excitation scheme into a microscopy modality. Previously in C-SFG microscopy, two parallel linearly polarized laser beams were coupled into an objective lens with a horizontal displacement in order to achieve non-collinear SPP scheme in a sample [13,14]. As shown in the inset in Figure 1 (b), this approach leaves more than half of the numerical aperture (NA) unused for each beam, thus increasing the focal spot size in both the lateral and the axial directions, which is determined by diffraction. The non-collinear setup also presents difficulties in alignment and integration with other microscope systems and components, such as confocal pin holes and scanning mirrors. In this letter, we propose and demonstrate a collinear C-SFG microscopy (CC-SFG) modality. As shown in Figure 1(c), two laser beams are aligned collinearly and fill the entire objective entrance aperture, delivering x and z polarized electric field to the sample respectively. As a result, the $\vec{P}^{NL}$ component along the y direction is generated via $\chi^{(2)}_{yxz}$, which could generate a propagating signal in the far field. The z polarized electric field at the focus can be realized by tightly focusing a radially polarized (RP) beam: a coherent vectorial beam with its polarization aligned along the radial direction. At the focal point, the lateral component of the field is cancelled out while the longitudinal component constructively interferes to form a highly confined z polarization state which can be utilized in various nonlinear imaging techniques including second and third harmonic generation, and coherent anti-Stokes Raman scattering [15]. Under a 1.4 NA focusing objective, maximal longitudinal electric field intensity at the focus could be larger than the transverse ones [16]. Furthermore, the spot size of longitudinal component was reported significantly smaller than that could be achieved using linear polarized beam [17], providing improved spatial resolution. As a result, RP beam has been applied in nanolocalization and nanostructure fabrication [18,19].

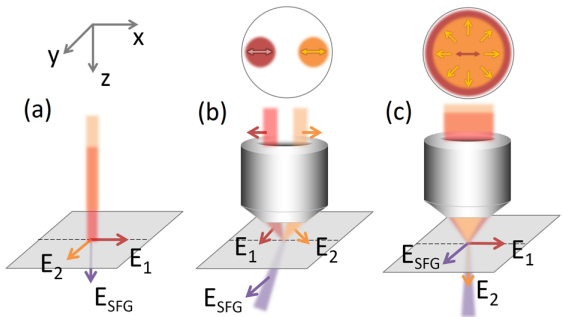

FIG. 1 (color online). Polarization configuration in C-SFG microscopy. $E_{SFG}, E_1, E_2$: Electric field of the SFG signal and two incident beams respectively. Insets in (b) and (c) show configurations of the excitation beam cross sections and polarization arrangement. (a) Collinear setup using linear polarized beam generates a longitudinally polarized C-SFG signal, which cannot effectively propagate in the forward direction in the far field. (b) Non-collinear setup using linear polarized beam can generate propagating C-SFG signal at the cost of reduced numerical aperture. (c) Collinear setup using linearly and radially polarized beams under tightly focused microscopy configuration can generate a propagating signal while utilizing the full numerical aperture.

Nonlinear wave mixing process under tightly focused scheme involves non-intuitive phase matching dependence. The complex focal fields of LP and RP beams are calculated respectively via vectorial diffraction theory [16]. In our simulation, two beams ($\lambda_{RP} = 520nm, \lambda_{LP} = 820nm$) are collinearly aligned with the optical axis of a 1.4 NA oil immersion objective, taking full use of the entrance aperture. Figure 2 (a) and (b) present the amplitude and phase distributions of the x-z plane of $E_{LPx}$, $E_{RPz}$ near the focus respectively, while Figure 2 (c) shows the Gouy phase [20] plotted along the optical axis for LP and RP beam respectively. Due the shorter wavelength and better confinement of longitudinal component [17], the RP beam yields a more tightly focused spot size and sharp transition of the Gouy phase. The inset of Figure 2(c) shows a π phase shift for LP beam, which is consistent with previous report [20]. Note that the second order nonlinear polarization is given by $P_{NL}^{(2)} = \chi_{yxz}^{(2)} E_{LPx} E_{RPz}$. We use the method in Ref.[21] to study the phase matching condition. Specifically, we consider a reference sphere centered at the focus and whose diameter (D) varies from 0.1μm to 6μm. For each diameter, we use Green's function to calculate the far field SFG signal generated by the $P_{NL}^{(2)}$ within the sphere; the signal power within an NA of 0.5 (to match the NA of the UV objective lens used in the experiment) is then integrated. The dependence of the integrated SFG signal on D is studied. Particularly, we consider both the scenario of focusing inside the bulk of uniform chiral liquid and the scenario of focusing on the interface between chiral liquid and achiral media. The monotonous increase of the C-SFG signal shown in Figure 2(d) indicates that in both bulk and interface regions the generated C-SFG signals are detectable; the signal generated at the interface becomes larger than that from the bulk for a sphere diameter > 3μm, due to the destructive interference effect arising from the Gouy phase change. This is consistent with previous report [13], where phase mismatching was greatly suppressed at the focal region with broken symmetry [21].

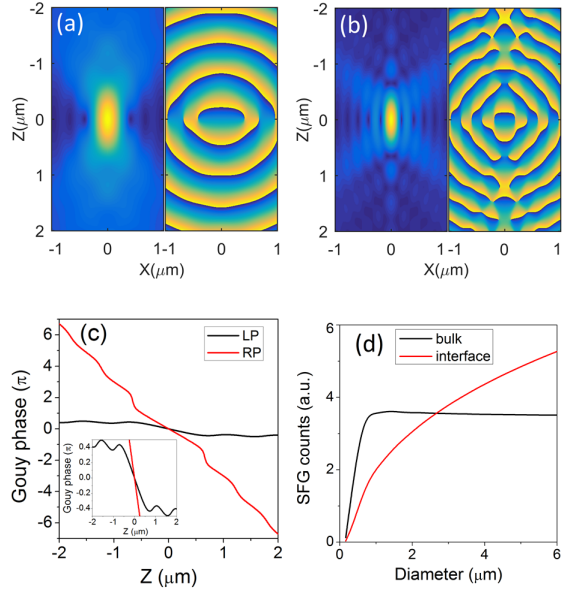

FIG. 2 (color online). Simulation results of CC-SFG microscopy. (a) Electric field amplitude and phase distributions near the focus of tightly focused linear polarized beam. (b) Electric field amplitude and phase distributions near the focus of tightly focused radially polarized beam. (c) Gouy phase plotting along the optical axis for LP (black) and RP (red) beam respectively, inset shows the enlarged plot of π phase shift from LP beam. (d) Far field C-SFG signal dependence of sample volume for signal generated at bulk (black) and interface (red) respectively.

In our experiments, laser pulses at 820nm was generated by a Ti:sapphire oscillator (Spectra-Physics, MaiTai) with 80 fs pulse width and 80 MHz repetition rate. The laser output was frequency-doubled into 410nm to pump an optical parametric oscillator (OPO, Spectra-Physics Inspire 100). The signal from the OPO was used as one of the pump beam ($\lambda_1 = 520nm$). The residual fundamental beam after the second harmonic generation was filtered by a long pass filter (ET542lp Chroma) and utilized as the second pump beam ($\lambda_2 = 820nm$). Therefore the wavelength of the SFG signal was 318.2nm. Note that due to the short wavelength a special ultraviolet objective was used to collect the generated SFG signal, which only had a NA of 0.5 as mentioned below. Both pump beams were collimated and expanded to 10 mm in diameter in order to fully use the NA of the excitation objective. The visible beam was converted into a radially polarized beam by a radial polarization converter (ARCoptix), and then transmitted through an uncoated pellicle beam splitter (Thorlabs, BP208) to combine with the x-polarized IR beam collinearly. An optical delay line in the RP beam path was used to fine adjust the temporal overlapping of the two beams. The combined collinearly propagating beam was directed into a home-made transmission-type microscope with a 1.4 NA focusing objective (Olympus, UPLSAPO 100XO) and a 0.5 NA collecting objective (Thorlabs, LMU-40X-UVB). Generated C-SFG signal passed through two filters (Thorlabs, FGUV11-UV; Chroma, ET313/25BP) and a polarizer (Thorlabs, GLB10-UV) before coupling into a spectrometer equipped with a liquid nitrogen cooled charge coupled device camera (Princeton Instrument, SP-2500, LN/400BR).

In order to demonstrate the imaging capability, a 5μm long cubic liquid cell was prepared on the surface of a fused silica slide by ion beam lithography as shown in Figure 3(a). 0.9 mol/L chiral R-BINOL solution was sandwiched between the slide and a cover glass, filling the liquid cell to form a chiral liquid cube in achiral environment as indicated in the inset. The generated sum frequency signal at 318.2nm coincides with the electronic resonance of BINOL molecule [22] and hence the signal can be significantly enhanced. The two pump beams were collinearly focused through the microscope into the liquid cell with an average power of 7mW at 820nm and 0.5mW at 520nm, respectively. The use of a low-power 520nm beam prevented two-photon excitation of fluorescent background from BINOL molecules.

By tightly focusing the RP and x-polarized LP incident beams, chiral SFG component of $\chi^{(2)}_{yxz}$ can be accessed by setting the detection polarizer in the y direction in the collinear microscopy. A typical measured C-SFG spectrum is shown in Figure 3(b), the peak wavelength matches well with predicted 318.2nm. Furthermore, when the sample was replaced by a racemic mixture of R-BINOL and S-BINOL solution with the same concentration, no signal was detected. It can be understood that $\chi^{(2)}_{yxz}$ for different enantiomers has opposite signs, and hence the nonlinear polarizations in the two enantiomer have a π phase difference, causing a total destructive interference and thus a vanishing signal as shown in Figure 3(b). The spectroscopic measurement indicates that the signal results from a chiral SFG process dominant by electric-dipole interaction; electric-quadrupole or magnetic-dipole interaction is negligible, giving rise to both chemical and chiral selectivity [11]. The microscopic imaging capability was demonstrated by raster scanning of the sample using a piezo stage at the sample plane located at 1μm away from the lower liquid-glass interface as shown in Figure 3(a). Obtained image of the liquid cell is presented in Figure 3(c), agreeing well with the scanning electron microscopy image in Figure 3(a).

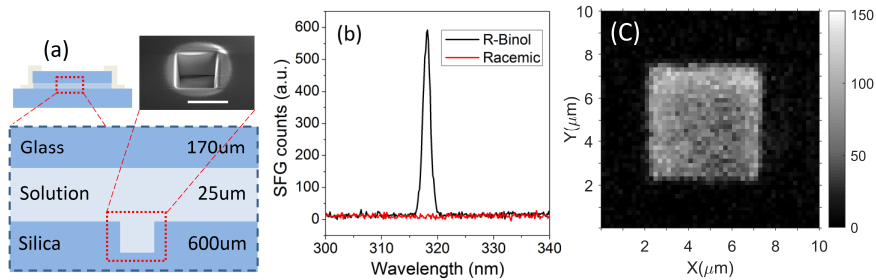

FIG. 3(color online). (a) SEM image of the cubic liquid cell, scale bar 5μm. (b) SFG spectrum from R-Binol solution and racemic mixture. (c) CC-SFG image of the cubic liquid cell.

To quantitatively characterize the spatial resolution, the transverse dimensions of the point spread functions (PSF) of both the CC-SFG and the non-collinear SFG microscopy were measured by using a knife edge method [23]. The same chiral liquid cell as shown in Figure 3(a) was used. For the non-collinear configuration, the two parallel pump beams were symmetrically displaced from the center of the objective lens by 1.6mm. Both beams were p-polarized to facilitate an SPP polarization arrangement [13]. Since C-SFG only occurs in the chiral liquid, the interface between the chiral liquid and the cell wall can be considered equivalent to a knife edge for the C-SFG characterizations. The interface was translated transversely across the focus within a 5μm range and at a 50nm step. The generated C-SFG intensity was recorded as a function of the knife edge positions. The measured distribution as shown in Figure 4 is essentially the cumulative curve of the PSF with respect to the wall traveling distance along either the x or y direction. In order to obtain the transverse dimensions of the PSF, numerical calculation based on the knife edge method was performed to fit the measured results. The intensity distributions used in the fitting were calculated by vectorial diffraction theory. Numerically fitted cumulative curves were obtained by using least-squares method according to experimental results, as plotted in Figure 4 (a)-(d) with the corresponding full width at half maximum (FWHM) of the PSF marked in the figures. Next, the lower interface of the chiral liquid cell was used for characterizing the longitudinal dimension of the PSF for both the collinear and the non-collinear configurations. Note that the interface produces enhanced signal due to the broken symmetry [13]. The measured FWHMs of longitudinal PSF are shown in Figure 4(e) and 4(f). Significantly improved spatial resolution on the y and z directions in the collinear C-SFG (~1.4 times enhancement) can be observed in comparison with the non-collinear arrangement from Figure 4. The collinear C-SFG also exhibits symmetric transverse PSF due to its center-symmetrical excitation scheme while the non-collinear C-SFG imaging exhibits an elliptically elongated transverse PSF distribution. It should be noticed that the effective NA used in the experiments did not reach the nominal maximum of 1.4 since all the imaging results were carried in a region deep inside the chiral liquid cell. The refractive index mismatch induced spherical aberration increases with the imaging depth [24] and results in a decreased effective NA. From the measured depth of field shown in Figure 4 (e), an effective NA of 0.75 can be inferred to result in a diffraction limited spatial resolution of ~420nm in x-y plane, agreeing well with our PSF measurements.

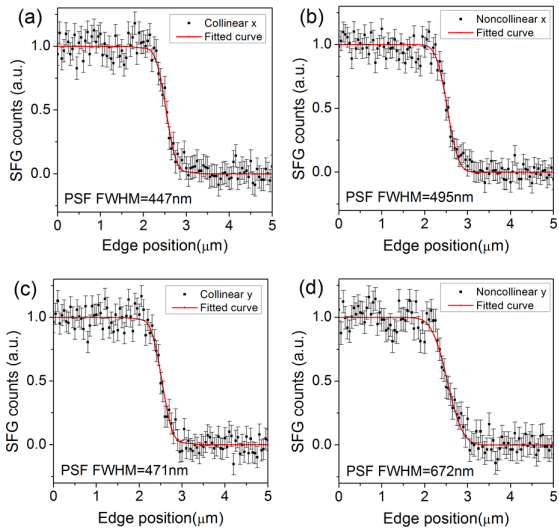

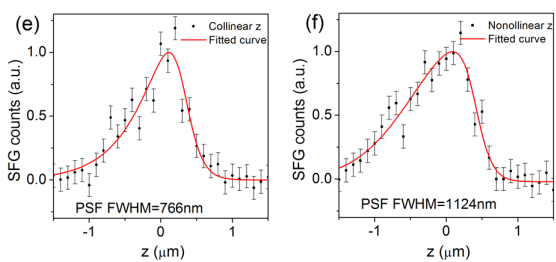

FIG. 4(color online). Measured dimensions of PSF for both the collinear and the non-collinear imaging arrangements. Cumulative distribution of the PSF (a) along the x direction in collinear C-SFG, (b) along the x direction in non-collinear C-SFG, (c) along the y direction in collinear C-SFG, (d) along y direction in non-collinear C-SFG, (e) along the z direction in collinear C-SFG, and (f) along the z direction in non-collinear C-SFG. The spatial resolutions of CC-SFG along the y and z directions are enhanced by ~1.4 times while the spatial resolutions along the x-direction are comparable. Error bars in the figures show the standard deviation.

In conclusion, we have demonstrated a chiral sum frequency generation microscopy based on longitudinally polarized laser field with collinearly and tightly focused excitation scheme. Collinear chiral SFG microscopy is demonstrated along with chiral spectroscopy capability and diffraction limited spatial resolution which conventional non-collinear SFG imaging could not reach. Using tightly focused radially polarized beam also has the potential for better excitation power efficiency as more than 75% of the incident power can be converted into longitudinally polarized components [16] under the excitation NA of 1.4. Based on the electric-dipole allowed chiral SFG contrast mechanism, the proposed technique can enable label-free, chemically/chiral selective SFG microscopy while allowing for ease of integration with existing microscope modalities. This technique can potentially lead to promising imaging developments for characterizing and spatially mapping chirality in biological and material systems.

KS acknowledges the funding support from the National Science Foundation of China (NSFC#11174019 and 61322509) and the Ministry of Science and Technology of China (National Basic Research Program of China under Grant No. 2013CB921904). The authors thank Chuanshan Tian in Fudan University for fruitful discussions.